\documentclass[10pt,conference]{IEEEtran}

\ifCLASSOPTIONcompsoc
  \usepackage[nocompress]{cite}
\else
  \usepackage{cite}
\fi

\usepackage{amssymb}
\usepackage{booktabs}
\usepackage{cite}
\usepackage{enumitem}
\usepackage[pdftex]{graphicx}
\usepackage{float}
\usepackage[framemethod=tikz]{mdframed}
\usepackage{tabularx}
\usepackage{tikz}
\usepackage{tipa}
\usepackage{xtab}
\usepackage[hidelinks]{hyperref}
\usepackage{bookmark}

\usepackage[]{todonotes}

\begin{document}

\setlist{nolistsep}

\title{Investigating the Interplay between Developers and Automation}
\author{
	\IEEEauthorblockN{Omar Elazhary}
	\IEEEauthorblockA{
		University of Victoria \\
		omazhary@uvic.ca
	}
}

\maketitle
\IEEEpeerreviewmaketitle

\begin{abstract}
Continuous practices are a staple of the modern software development workflow.
Automation, in particular, is widely adopted due to its benefits related to quality and productivity.
However, automation, similarly to all other aspects of the software development workflow, interacts with humans (in this case developers).
While some work has investigated the impact of automation on developers, it is not clear to what extent context and process influence that impact.
We present our ADEPT theory of developers and automation, in an attempt to bridge this gap and identify the possible ways context, process, and other factors may influence how developers perceive, interpret, and interact with automation.
\end{abstract}

\begin{IEEEkeywords}
Software engineering, automation, continuous software development, continuous integration, software engineering theory.
\end{IEEEkeywords}


%

\begin{figure*}[h!t]
	\centering
	\includegraphics[width=1.7\columnwidth]{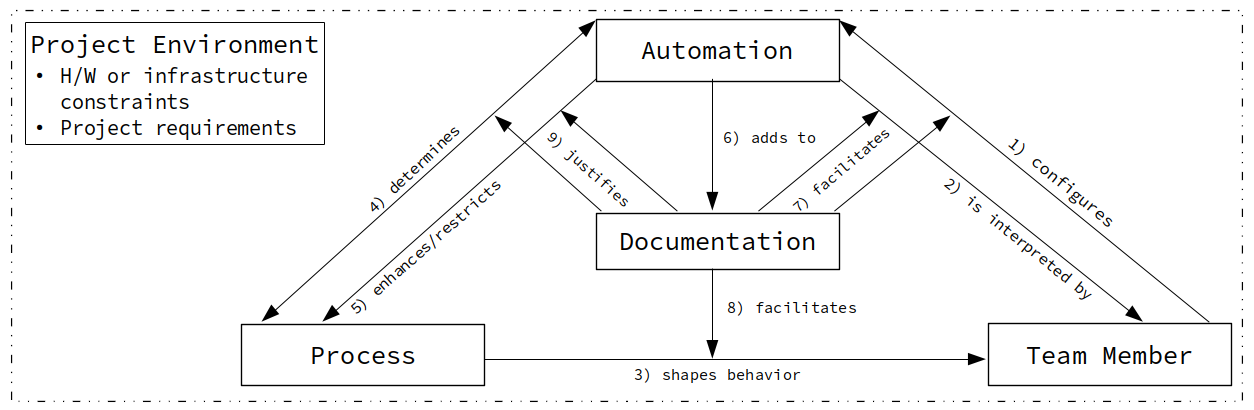}
	\vspace{-4mm}
	\caption{Visual Representation of the ADEPT Theory (Boxes are constructs, Edges are relationships)}
	\label{fig:theory}
	\vspace{-5mm}
\end{figure*}
\section{Introduction}
\label{sec:intro}

Continuous practices are a staple of modern software development \cite{fitzgerald2017continuous}.
The most commonly adopted continuous practice is automating some tasks of the development workflow.
These tasks typically include building an application, testing it, releasing it, and possibly deploying it to a client's system.
Given the benefits to productivity and quality such automation brings \cite{vasilescu2015quality, hilton2016usage}, it is no wonder such tools have seen widespread adoption.

However, automation in general has had an uneasy relationship with humans \cite{parasuraman1993performance, parasuraman2010complacency, singh1993automation, parasuraman1997humans}.
One phenomenon related to automation is that of misuse, or as it is later called complacency, which denotes the over-reliance on the results produced by automation.
The opposite extreme to that phenomenon is disuse which describes a lack of trust in the results produced by automation.
Finally, there is automation abuse, which captures automation that is configured in such a way that it hinders the process it is meant to improve.

In the software engineering domain, particularly in the context of continuous practices, such phenomena also occur in addition to new phenomena specific to software engineering.
For instance, software developers can be impacted by automated build results, which affects their next commit \cite{souza2017sentiment}.
Or the negative impact adopting automated build tools has on a project's developer-base \cite{gupta2017impact}.

Unfortunately, the interplay between humans and automated tools has been under-explored in software engineering.
We hope to explore this area and map out the various phenomena that relate to the interactions between software developers and the automation they use in their workflows.
\section{Background and Related Work}

Most of the literature regarding automation in the context of continuous practices is centered around two major areas of research: optimizing the automation workflow, and detecting problems with automation configurations.
Optimization efforts typically center around selectively or incrementally \cite{maudoux2017bringing} executing tests or builds, or exploring how to shorten build durations \cite{ghaleb2019empirical}.
Detecting problems with configurations, on the other hand, focuses on documenting different incorrect configuration methods \cite{gallaba2018use, zampetti2020empirical}, or designing tools that detect such incorrect configurations \cite{vassallo2019automated}.

More recently, however, there has been some work on the human aspects of continuous practices.
As mentioned in section \ref{sec:intro}, some studies investigated the relationship between failing builds and developer sentiment over time \cite{souza2017sentiment}.
Souza and Silva observe that failing builds relate to negative commit messages when trying to fix the error.
And negative commit messages are more likely to produce failing builds.
Another example of automation having an impact on human behavior is the work done by Gupta et al. \cite{gupta2017impact}, where they observe that a project's ability to retain and attract developers decreases upon adopting Travis CI, a popular continuous integration tool.
However, it is not clear \emph{why} these phenomena occur.

\section{Previous Studies}

To map out the different relationships between developers and automation, we started by examining open source projects and how automation was used within that context.
We examined the contribution guidelines for several open source projects to examine how newcomers were being instructed to use and interpret the automation that was attached to the project repositories \cite{elazhary2019not}.
Surprisingly, we found that none of the projects discussed their attached automation tools beyond them being used for testing purposes.

We then decided to investigate continuous practices in an industrial context, and how they might impact development workflow.
To that end, we conducted two studies; the first explored the impact of continuous practices on non-functional requirements \cite{werner2020continuous}, and the second focused on the perceived benefits and costs of continuous practices \cite{elazhary2020tse}.
In the former study, we found that the automation brought about by continuous practices had facilitated dealing with particular non-functional requirements (such as configurability) with developers reprioritizing how they deal with these requirements.
In the second study, we observed that the benefits and costs brought about by continuous practices (including automation) are a result of the practices, as opposed to the overarching practice set.
For instance, automation was perceived to increase development cycle velocity, but--without tests--it had no impact on quality.
The association of costs and benefits with their finer-grained practices (as opposed to the overarching methodology of continuous integration) helped us better map the relationship between developers and automation.
The mapping, in turn, led to our most recent study.
\section{Proposed Solution and Results}

To gain a better understanding of how developers interact with automation in their specific contexts, we decided to develop a theory.
We considered this a step towards a more theory-oriented software engineering research field as argued by Stol and Fitzgerald \cite{stol2015theory}.
To the best of our knowledge at the time, there was no theory explaining the different ways developers interacted with automation.

Based on our previous studies \cite{elazhary2019not, elazhary2020tse}, we attempted to create an explanatory theory to better represent the phenomena we were observing and to provide a starting point to generate hypotheses about human-centric phenomena related to automation \cite{elazhary2021nier}.
The ADEPT theory (\textbf{A}utomation, \textbf{D}ocumentation, Project \textbf{E}nvironment, \textbf{P}rocess, \textbf{T}eam Member) maps the relationships we observed and sets the scene for further investigation.
A representation of ADEPT can be seen in Figure \ref{fig:theory}.

\textbf{Team member} refers to people who interact with automation, be they developers, program managers, testers, or others.
\textbf{Automation} is abstracted to include automated tools that team members interact with throughout their development process.
\textbf{Process} encapsulates the various practices and activities that constitute their development workflow.
\textbf{Documentation} represents the artifacts with which developers communicate knowledge about their process, tools, and code.

The edges between the constructs represent relationships we observed during our previous studies \cite{elazhary2020tse}.
Two relationships stand out in particular; automation adds to documentation, and the bidirectional relationship between automation and process.
We found that automation serves to document tacit build knowledge in form of configuration and build scripts.
We also found that the desire to implement a particular process impacts the choice of automation used.
We also hypothesize that the reverse is also true in that the desire to use a particular tool can force developers to implement a different process.

The theory captures context as available through both the project environment and the process within which developers operate.
In fact, it illustrates that these relationships may only produce the results discussed by literature under a constant context, which is not the case.
We also found that the relationship between developers and automation is impacted by documentation.
In many cases, developers would resort to a tool's documentation to either run an automated build manually, or interpret a tool's results.

Furthermore, our theory allows for the formation of compound relationships.
For instance, while a developer can use documentation to interpret the results of automation, that behavior is shaped by the process within which they operate.
Without the process priming that particular behavior, developers could interpret automation results differently.
\section{Future Work}

While our previous studies provided us with an understanding of possible relationships between developers and automation, we need to operationalize the concepts our theory proposes.
And we still do not know if these relationships apply in different contexts.
We make the argument that context is important, and the results should always be interpreted within it, however, operationalizing the context is not as straightforward as it seems.

Currently, we plan to investigate how best to operationalize these concepts, and test the relationships we posit in our theory.
We plan to investigate if the relationships can be used to interpret previous results in both industry and open source contexts to contrast these differences.
The main objective is to determine to what extent our theory describes the relationship between automation and developers, and whether it can be improved.

Similarly to our previous studies, we plan to use a mixed-methods approach to data collection.
We plan to conduct online surveys that help us capture context and human-centric aspects of the relationships we observed.
Additionally, we plan to mine repository development logs to both triangulate the survey results, and objectively observe development metrics and relate them to automation properties.

\ifCLASSOPTIONcaptionsoff
  \newpage
\fi

\bibliographystyle{IEEEtran}
\bibliography{main}

\end{document}